\documentclass[conference]{IEEEtran}
\IEEEoverridecommandlockouts
\usepackage{cite}
\usepackage{amsmath,amssymb,amsfonts}
\usepackage{algorithmic}
\usepackage{graphicx}
\usepackage{textcomp}
\usepackage{xcolor}
\usepackage{multirow}
\usepackage{array}
\usepackage{subfig}
\usepackage[linesnumbered,ruled,vlined]{algorithm2e}
\usepackage{comment}
\usepackage{soul}
\usepackage{url}
\def\BibTeX{{\rm B\kern-.05em{\sc i\kern-.025em b}\kern-.08em
    T\kern-.1667em\lower.7ex\hbox{E}\kern-.125emX}}

\begin{document}

\title{A Zero-Knowledge Signature Framework for Efficient Post-Quantum Message Authentication in Cooperative Automated Driving*\\ 
{\footnotesize \textsuperscript{*}Note: This paper is an extended version of a conference paper titled: ``ZKS-PQC: Zero Knowledge Signature for PQC''}
\thanks{
This work was supported by CyberSecurity Research Flanders with reference number VR20192203.}
}
\author{
	\IEEEauthorblockN{Takahito Yoshizawa\IEEEauthorrefmark{1}, Aysajan Abidin\IEEEauthorrefmark{1}, Edoardo Peña-González\IEEEauthorrefmark{1}, Bart Preneel\IEEEauthorrefmark{1}} \\
	\IEEEauthorblockA{\IEEEauthorrefmark{1}COSIC, KU Leuven, \\
		Kasteelpark Arenberg 10 Bus 2452, B-3001 Leuven, BELGIUM
		\\\{takahito.yoshizawa, aabidin, edoardo.pena-gonzalez, bart.preneel\}@esat.kuleuven.be}
}

\maketitle

\begin{abstract}
 Connected and Automated Vehicles (CAV) rely on authenticated Vehicle-to-Everything (V2X) communications to exchange safety-critical information among vehicles and roadside infrastructure. 
 As the automotive industry transitions toward post-quantum cryptography (PQC), the significantly larger public keys and signatures of standardized PQC digital signature algorithms introduce substantial communication overhead, which challenges the scalability of certificate-based V2X authentication, particularly for high-frequency cooperative awareness messages (CAM). 
 This paper presents ZKS-PQC, a \textit{zero-knowledge signature framework} that enables communication-efficient post-quantum message authentication for cooperative V2X systems. 
 Instead of transmitting complete post-quantum public keys and signatures, the proposed framework replaces this authentication material with a compact Zero-Knowledge Proof (ZKP) while preserving compatibility with existing certificate-based trust architectures. 
 This approach supports incremental migration and backward compatibility with the legacy ECDSA. 
 The framework is implemented using the Open Quantum Safe (liboqs) and ZKP (Bulletproofs) libraries, and we evaluated the performance of standardized NIST PQC signature algorithms and additional candidate algorithms. 
 Experimental validation on both a Linux platform and a commercial On-Board Unit (OBU) demonstrates substantial reductions in message size exceeding 95\% for all algorithms, while limiting additional processing overhead by staying within the order of milliseconds at both sender and receiver in many algorithms, consistent with the latency requirements of real-time V2X operation. 
 By decoupling communication overhead from the size characteristics of post-quantum signature algorithms, ZKS-PQC offers a practical migration strategy for scalable, quantum-resilient message authentication in future CAV systems.\footnote{This paper is an extended version of the conference paper titled: ``ZKS-PQC: Zero Knowledge Signature for PQC.'' The original 6-page paper is expanded to twice the length by adding a significant amount of new content. Details of the changes are described in Sect.~\ref{sec:contribution}.}
\end{abstract}


\section{Introduction} \label{sec:intro}
In order to stay ahead of the evolution of quantum computers, the National Institute of Standards and Technology (NIST) launched the standardization process in 2016 to select Post-Quantum Cryptography (PQC) algorithms.
This includes public-Key Encapsulation Mechanism (KEM) and Digital Signature Algorithms (DSA).
After three rounds of analysis and evaluation, NIST standardized one KEM and three DSAs in 2022~\cite{nistIr8413upd1}.\footnote{The standardized KEM algorithm is CRYSTALS–KYBER (ML-KEM (FIPS 203)), and the three standardized DSAs are CRYSTALS–Dilithium (ML-DSA (FIPS 204)), FALCON (FN-DSA (FIPS 206)), and SPHINCS+ (SLH-DSA (FIPS 205). Both Dilithium and Falcon are lattice-based and SPHINCS+ is a hash-based algorithm.} 
Following this, in order to diversity the DSA portfolio, NIST published a call for additional signature schemes~\cite{nistAdditional2023}.
Initially, there were 40 candidate algorithms. 
The number is narrowed down to 14 candidates at the end of the subsequent first round evaluation~\cite{nistIr8528} and further down to 9 candidate in the second round evaluation~\cite{nistIr8610}. 
This evaluation process is ongoing at the time of this writing~\cite{nistpqcurl}.

Along with the PQC algorithm selection process, NIST published a expected migration timeline toward PQC algorithms in 2024~\cite{nistIr8547}.
Other countries and regions, such as EU, UK, Canada, and Australia, also established a similar timeline~\cite{el2026institutional}. 
Overall, 2035 is the commonly shared timeline to deprecate the existing algorithms and complete the migration toward PQC algorithms.
Specifically, public-key cryptography algorithms such as RSA and elliptic curve (EC)-based key exchange and signature schemes are vulnerable because their security relies on computationally hard problems, including large-integer factorization and the discrete logarithm problem.

\subsection{Characteristics of PQC Signature Algorithms}
The key characteristics of the PQC DSAs are: (1) either the public key or signature length is significantly longer than that of ECDSA, and (2) some algorithms are significantly slower than ECDSA.\footnote{The exact numbers of the public key/signature size and processing time vary from one algorithm to another.} 
These characteristics will impact systems that use communication channels or real-time systems. 

The communication channel is affected as a longer public key or signature makes the message longer, making a single message occupy the channel for a longer period.
Furthermore, messages longer than a frame size require message fragmentation.
This reduces the effective channel capacity and incurs additional processing in both senders and receivers. 
Extra processing time for signature generation and verification will delay the end-to-end message latency. 
An excessive end-to-end latency can potentially compromise the integrity of the real-time system (RTS) or render it useless.
Finally, resource-constrained IoT devices may not be able to support PQC algorithms due to their limited storage capacity or computing power.

The conventional messages include both a certificate (containing a public key) and a signature. 
This way, receivers of these messages can verify the signature in the message.
If we apply this approach with the PQC DSAs, the message size can become excessively long, stressing the communication channel, processing capacity in both transmitters and receivers, and causing excessive delay.
This situation limits the usable algorithms to those that have shorter public keys and signatures.
However, it artificially limits the pool of available PQC algorithms.
This is far from an ideal situation; ideally, algorithm selection should be based on cryptographic properties and the desired security level for the intended application and operational circumstances, rather than being limited by technical constraints. 
In light of this situation, a new approach is necessary to eliminate, or at least minimize, these impacts in systems vulnerable to adopting PQC DSAs. 

\subsection{Impact of PQC Migration on V2X Communication}
Vehicular communication (V2X) fits the characteristics described above: an increase in communication channel usage, additional latency in real-time processing in a resource-constrained operational environment.
In particular, spectrum allocation for Cooperative Intelligent Transport Systems (C-ITS) is limited to 20 MHz bandwidth in many countries.
Previous work has shown that, even in the best-case scenario (Falcon-512), the channel capacity will decrease by half when approximately 30\% of vehicles on the road support a hybrid mode of operation~\cite{yoshizawa2023post}.\footnote{\textit{Hybrid mode} means that both conventional ECDSA and a new PQC signature algorithm are supported at the same time by digital certificates to hold both ECDSA public key and PQC public key, and ECDSA signature and PQC signature are included in the V2X messages.} 

This situation will saturate the communication channel, making the whole V2X communication untenable.
This problem is particularly significant for Cooperative Awareness Messages (CAM), which are broadcast at frequencies of up to 10 Hz~\cite{etsien3026372v1412019}.
Furthermore, vehicle on-board units (OBUs) with limited processing and storage capacity may not be able to store a large number of large public keys. 
Additionally, signature generation and verification may exceed the system’s hard real-time constraints, rendering real-time processing infeasible.
These limitations have significant implications for the integrity of V2X communication and its intended goal of improving road safety.

\subsection{CAV-Related Regulatory Background}
Connected and automated vehicles (CAV) deployments are increasingly dependent on communication with other vehicles, infrastructure, and backend services. 
In this regard, the United Nations (UN) recently adopted the first internationally harmonized Whole-Vehicle Type-Approval framework (WVTA) for fully Automated Driving Systems (ADS)~\cite{UNECEadopts2026,UNECEproposal2026a,UNECEproposal2026b}. 
Given that EU member states collectively constitute a substantial share of the contracting parties participating in the UN type-approval regulation, the EU coordinated risk assessment on connected and automated vehicles provides a relevant regional basis for examining the cybersecurity risks associated with ADS deployment~\cite{NISEUcoordinated2026}. 
The assessment identifies communication and connectivity systems, cloud and backend systems, vehicle-control systems, and supply chains among the relevant cybersecurity risk areas, making trustworthy connectivity an enabling condition for deployment~\cite{NISEUcoordinated2026}. 
Under WVTA, ADS shall be protected from cyber threats, with compliance demonstrated through UN Regulation No. 155,~\cite{UNECEproposal2026a,UNECEreg1552021}. 
At the V2X level, said regulation indicates mitigation of threats related to ``vehicle communication channels'', and ETSI TS 103 097 defines the certificate and message-security mechanisms based on digital certificates and signatures (i.e., with the use of the Elliptic Curve Digital Signature Algorithm (ECDSA))~\cite{etsiTs1030972022}. 

Against this backdrop, cryptographically relevant quantum computers will undermine current signature schemes such as ECDSA. 
NIST’s transition plan proposes deprecating 112-bit ECDSA after 2030 and disallowing ECDSA at all approved security strengths after 2035~\cite{nistIr8547}. 
Because the mitigation of threats such as ``malicious V2X messages'' or ``spoofing'' in UN Regulation No. 155 is the vehicle’s verification of authenticity and integrity of incoming messages, certificate-based V2X trust architectures must migrate toward post-quantum digital signatures to uphold the regulation cybersecurity requirements.

\subsection{Cooperative V2X Communication in ADS}
The EU Coordinated risk assessment identifies communication and connectivity systems as critical CAV assets interconnected with vehicle control and backend systems~\cite{NISEUcoordinated2026}. 
Within this system context, V2X enables CAV to exchange operational information with other vehicles and roadside infrastructure.

Cooperative Awareness Messages (CAM) contain vehicle-state information, including position, speed, and heading~\cite{etsien3026372v1412019}. 
Because received CAM may influence safety-relevant vehicle behavior, their authenticity and integrity must be verified before use. 
Farrell et al. (2024) showed that sensor fusion can improve robustness against spoofed CAM messages. 
However, that security analysis must still consider both message-level threats and CAV behavior following the message reception because an attacker may combine delayed CAM verification with spoofing or blinding of the sensor stack (e.g., LiDAR, cameras)~\cite{farrell2023security}.

For vehicle ITS stations, ETSI EN 302 637-2 specifies CAM generation intervals from 100 to 1,000 msec, corresponding to rates from 10 to 1 Hz, with the actual rate governed by vehicle dynamics and channel congestion~\cite{etsien3026372v1412019}. 
Because every vehicle on the road broadcasts CAM repeatedly, CAM traffic presents a demanding authentication workload. 
Authentication-material size affects channel usage and scalability, while verification latency affects whether cooperative information is available in time for safety-relevant functions.
This high-rate certificate-authentication workload provides the application context for the post-quantum V2X migration approaches reviewed in Sect.~\ref{sec:relatedwork}.

\subsection{Our Contribution} \label{sec:contribution}
In this paper, we introduce a new approach to remove these constraints.
Our approach neutralizes the stated impacts and broadens the choice of usable PQC DSAs, not limited to the smaller size signature and public keys only.
Our experiment demonstrates that our scheme reduces the message size by more than 95\% compared to the conventional approach in majority of the PQC signature algorithms.  
As a result, our scheme removes the obstacles to using any PQC DSAs with any desired security level.

This paper is an extended version of what we previously submitted to a conference.
We have expanded the content by: (1) adding new simulation results using a dedicated V2X hardware platform, thus significantly expanding the scope of our experiments (Sect.V), (2) significantly expanding tables to discuss simulation results in numerical values to present additional details (Table 2, 3), (3) introducing supplemental figures, tables, and texts in various parts of the paper to supplement texts to enhance readability (Fig.1, Table 1, etc.), (4) further introducing new subsections on CAV/ADS aspects (Sect.I.C, D, Sect.V.D), and (5) rewriting the abstract to position the paper on CAV/ADS aspects. 
As a result, this extended version has twice the length of the previous paper with wider and more detailed coverage of related aspects. 
With the introduction of this additional content, we have also changed the paper title to reflect our focus on CAV, ADS, and V2X implications of the PQC migration.

This paper is organized as follows. 
We first review related work in Sect.~\ref{sec:relatedwork}, followed by giving a brief overview of Zero Knowledge Proof (ZKP) in Sect.~\ref{sec:zkpprimer}. 
We then present our scheme in Sect.~\ref{sec:proposed_scheme}, discuss our experiment and its results in Sect.~\ref{sec:experiment}. 
Sect.~\ref{sec:conclusion} concludes our discussion.

\section{Related Work} \label{sec:relatedwork}
There are several previous works that discuss PQC migration, and some of them specifically focus on issues and solutions in the context of V2X communication. 
We briefly cover these works.

A study in PQC migration on C-ITS explored a possible avenue toward PQC adoption with a minimum impact on the co-existence of legacy and PQC algorithms~\cite{lonc2023feasibility}.
It suggested that the best practical solution is to modify the existing PKI certificate-based standards. 
Another study covered the usage of cryptography in vehicles and identifies areas that require PQC algorithm support~\cite{lohmiller2025survey}. 
This included vehicle-internal network, OEM-specific communication, and V2X communication.
This paper identified research gaps and proposed possible solutions. 


To realize gradual and orderly migration from the conventional ECDSA to PQC DSA, both algorithms need to co-exist at the same time. 
For this purpose, a so-called \textit{hybrid certificate} was proposed~\cite{chen2024hybrid}.
In this scheme, public keys of both algorithms are carried in the certificate, and messages contain this hybrid certificate as well as the message signatures of both algorithms. 
This increases the overall message size, and some of the standardized PQC DSAs exceed the frame size, requiring fragmentation.
In a subsequent study, the use of a hybrid certificate was shown to saturate the communication channel as the proportion of vehicles that support a hybrid certificate increases~\cite{yoshizawa2023post}.

Several studies propose approaches to improve efficiency in light of message fragmentation incurred by the use of PQC DSA in V2X communication.
First, a proposed split-signature scheme addresses the fragmentation of signatures in multiple frames~\cite{kim2023signature}. 
This scheme allows independent per-frame verification of fragmented signatures. 
Second, a hybrid design of PQC authentication is presented~\cite{bindel2022drive}.
Its goal is to maintain efficiency and security across multiple fragmented message frames.
Its three solution schemes have a varying level of effectiveness while trying to maintain backward compatibility. 
The discussion is theoretical in nature, and there is no proof-of-concept implementation or quantitative evaluation of the proposed schemes.
Third, an efficient certificate transmission scheme is proposed in safety messages~\cite{twardokus2024cryptography}.
It introduces a \textit{partially hybrid certificate} scheme, in which the public key in the certificate is signed by both ECDSA and PQC algorithms. 
It also suggests an enhancement to reduce redundant transmission of certificates to improve the channel usage over the existing scheme in the standards. 

The common issue across these proposed schemes is that the fundamental need to verify the integrity of the entire message as a whole is still required across the set of fragmented frames.
Clearly, the best situation is to avoid message fragmentation altogether, implying limiting the message size increase within the maximum frame length. 
In this sense, the level of performance improvement in these schemes - given that the fragmentation is unavoidable - is rather limited.


In summary, previous related work falls under one of these types: (1) analyzes the impacts and implications of PQC migration in C-ITS at the theoretical level, and (2) proposes alternative schemes to mitigate the migration impacts.  

\section{ZKP Primer} \label{sec:zkpprimer}

\begin{table*}[]
	\centering
	\caption{ZKP Protocols Overview}
	\label{tab:zkp_overview}
	\begin{center}
		\begin{tabular}{p{2.5cm}p{2cm}p{2.5cm}p{2cm}p{2cm}}
			\noalign{\hrule height 1pt}
			\multicolumn{1}{c}{\textbf{Protocol Name}} & 
			\multicolumn{1}{c}{\textbf{Protocol Family}} & 
			\multicolumn{1}{c}{\textbf{Interactive vs. Non-interactive}} & 
			\multicolumn{1}{c}{\textbf{Trusted Setup vs. Transparent}} & 
			\multicolumn{1}{c}{\textbf{Single vs. Universal}} \\ 
			\noalign{\hrule height 1pt}
			
			\multicolumn{5}{c}{Pairing SNARKs (QAP)} \\ \hline
			\multicolumn{1}{c}{Pinocchio}        & 
			\multicolumn{1}{c}{zk-SNARK}          & 
			\multicolumn{1}{c}{Non-interactive (FS)}        & 
			\multicolumn{1}{c}{Trusted setup}          & 
			\multicolumn{1}{c}{Single}        \\ 
			\multicolumn{1}{c}{Geppetto}        & 
			\multicolumn{1}{c}{zk-SNARK}        & 
			\multicolumn{1}{c}{Non-interactive (FS)}          & 
			\multicolumn{1}{c}{Trusted setup}        & 
			\multicolumn{1}{c}{Single}       \\ 
			
			\multicolumn{1}{c}{Groth-16}        & 
			\multicolumn{1}{c}{zk-SNARK}          & 
			\multicolumn{1}{c}{Non-interactive}        & 
			\multicolumn{1}{c}{Trusted setup}          & 
			\multicolumn{1}{c}{Single}        \\ 
			\noalign{\hrule height 1pt}
			
			\multicolumn{5}{c}{Universal SNARKs} \\ \hline
			\multicolumn{1}{c}{Sonic}        & 
			\multicolumn{1}{c}{zk-SNARK}          & 
			\multicolumn{1}{c}{Non-interactive (FS)}        & 
			\multicolumn{1}{c}{Trusted setup}          & 
			\multicolumn{1}{c}{Universal}        \\ 
			\multicolumn{1}{c}{Marlin}        & 
			\multicolumn{1}{c}{zk-SNARK}          & 
			\multicolumn{1}{c}{Non-interactive (FS)}        & 
			\multicolumn{1}{c}{Trusted setup}          & 
			\multicolumn{1}{c}{Universal}        \\ 
			\multicolumn{1}{c}{PLONK}        & 
			\multicolumn{1}{c}{zk-SNARK}          & 
			\multicolumn{1}{c}{Non-interactive (FS)}        & 
			\multicolumn{1}{c}{Trusted setup}          & 
			\multicolumn{1}{c}{Universal}        \\ 
			\noalign{\hrule height 1pt}
			
			\multicolumn{5}{c}{Transparent SNARKs} \\ \hline
			\multicolumn{1}{c}{Aurora}           & 
			\multicolumn{1}{c}{zk-SNARK}           & 
			\multicolumn{1}{c}{Interactive → FS}           & 
			\multicolumn{1}{c}{Transparent}           &  
			\multicolumn{1}{c}{Universal}		   \\ 
			\multicolumn{1}{c}{STARK}           & 
			\multicolumn{1}{c}{zk-STARK}           & 
			\multicolumn{1}{c}{Interactive oracle proof → FS}           & 
			\multicolumn{1}{c}{Transparent}           &  
			\multicolumn{1}{c}{Universal}		   \\ 		
			\multicolumn{1}{c}{SuperSonic}           & 
			\multicolumn{1}{c}{zk-SNARK}           & 
			\multicolumn{1}{c}{Non-interactive (FS)}           & 
			\multicolumn{1}{c}{Transparent}           &  
			\multicolumn{1}{c}{Universal}		   \\ 
			\noalign{\hrule height 1pt}
			
			\multicolumn{5}{c}{Inner-Product Argument (IPA) systems} \\ \hline
			\multicolumn{1}{c}{Bulletproofs}           & 
			\multicolumn{1}{c}{Bulletproofs}           & 
			\multicolumn{1}{c}{Non-interactive}           & 
			\multicolumn{1}{c}{Transparent}           &  
			\multicolumn{1}{c}{Universal}		   \\ 
			\multicolumn{1}{c}{Halo}           & 
			\multicolumn{1}{c}{zk-SNARK}           & 
			\multicolumn{1}{c}{Non-interactive (FS)}           & 
			\multicolumn{1}{c}{Transparent}           &  
			\multicolumn{1}{c}{Universal}		   \\ 
			\noalign{\hrule height 1pt}
			
			\multicolumn{5}{c}{MPC-in-the-head} \\ \hline
			\multicolumn{1}{c}{Ligero}           & 
			\multicolumn{1}{c}{zk-SNARK}           & 
			\multicolumn{1}{c}{Interactive → FS}           & 
			\multicolumn{1}{c}{Transparent}           &  
			\multicolumn{1}{c}{Universal}		   \\ 
			\multicolumn{1}{c}{Zilch}           & 
			\multicolumn{1}{c}{zk-STARK}           & 
			\multicolumn{1}{c}{Interactive → FS}           & 
			\multicolumn{1}{c}{Transparent}           &  
			\multicolumn{1}{c}{Universal}		   \\ 
			\noalign{\hrule height 1pt}
			
			\multicolumn{5}{c}{Goldwasser-Kalai-Rothblum (GKR) Protocol / sumcheck} \\ \hline
			\multicolumn{1}{c}{Virgo}           & 
			\multicolumn{1}{c}{zk-SNARK}           & 
			\multicolumn{1}{c}{Interactive → FS}           & 
			\multicolumn{1}{c}{Transparent}           &  
			\multicolumn{1}{c}{Universal}		   \\ 
			\multicolumn{1}{c}{Hyrax}           & 
			\multicolumn{1}{c}{zk-SNARK}           & 
			\multicolumn{1}{c}{Non-interactive (FS)}           & 
			\multicolumn{1}{c}{Transparent}           &  
			\multicolumn{1}{c}{Universal}		   \\ 
			\noalign{\hrule height 1pt}
		\end{tabular}
	\end{center}
\end{table*}

In this section, we give a brief overview of ZKP.\footnote{For further details of ZKP protocols, interested readers are encouraged to read further sources such as~\cite{sheybani2025zeroknowledgeproofframeworkssystematic} and \cite{motlagh2025technology}.}
The ZKP concept involves two entities: a \textit{prover} and a \textit{verifier}, in which the former proves knowledge (e.g., possession of secret information) to the latter without revealing that knowledge itself.
ZKP uses a mathematical construct to hide knowledge in such a way that the \textit{verifier} concludes that the \textit{prover} indeed has knowledge based on the presented proof.
It uses the property of a one-way function so that it is easy for the \textit{prover} to calculate a proof and the \textit{verifier} to verify it, but difficult for entities other than the \textit{prover} to generate a valid proof.
The underlying mathematical problem includes ones such as the Schnorr algorithm~\cite{schnorr1989efficient}. 
Because the \textit{prover} and \textit{verifier} directly communicate in ZKP, it does not require any trusted third party. 
Due to this decentralized nature, one of the main applications is the blockchain, which is the enabling technology in cryptocurrency~\cite{guan2020blockmaze}.

ZKP has the following properties~\cite{wikiarticlezkp}:
\begin{enumerate}
	\item Completeness - if the statement is true, an honest prover will convince an honest verifier.
	\item Zero-knowledge - if the statement is true, then the verifier will learn nothing more than the fact that the statement is true, i.e., the prover knows the underlying secret.
	\item Soundness - if the statement is false, then no dishonest prover can convince an honest verifier.
\end{enumerate}

There are three types of ZKP protocols: (1) Zero-Knowledge Succinct Non-interactive ARgument of Knowledge (zk-SNARK) in 2012~\cite{bitansky2012extractable}, (2) Bulletproofs in 2017~\cite{bunz2018bulletproofs}, and (3) Zero-Knowledge Scalable Transparent ARgument of Knowledge (zk-STARK) in 2018~
\cite{ben2018scalable}.\footnote{It should be noted that zk-SNARK refers to a family of protocols rather than a single protocol, while zk-STARK refers to a specific protocol. Due to this reason, zk-SNARK protocols are collectively referred as zk-SNARKs~\cite{wikiarticlezkpprotocol}.}
These protocols are characterized by these properties:
\begin{itemize}
	\item \textit{Succinctness}: the proof is a fixed size regardless of the size of the secret information. One of the main characteristics of zk-SNARKs is this property. 

	\item \textit{Interactive} vs. \textit{non-interactive}: the former requires message exchange between the \textit{prover} and the \textit{verifier}. It is a 3-way interaction of messages called: \textit{commitment}, \textit{challenge}, and \textit{response}. On the other hand, the latter is a single one-way message from the \textit{prover} to the \textit{verifier} -- a \textit{prover} sends both a proof and a commitment within a single message to the \textit{verifier}, who verifies the proof based on the received information without further communication with the \textit{prover}.\footnote{The first ZKP protocols were interactive type. However, the difference between them has been blurring due to the use of Fiat-Shamier heuristic~\cite{fiat1986prove},  which replaces the verifier's random \textit{challenge} value with a hash of commitment and statement (i.e., $hash(c,s)$, where $c$ and $s$ are a commitment and a statement, respectively). This way, an explicit \textit{challenge} message from the verifier is no longer needed, removing the interactive message exchange.}

	\item \textit{Trusted setup} vs. \textit{transparent}: a \textit{trusted setup} is an initialization procedure between the \textit{prover} and the \textit{verifier}. This setup establishes public parameters (called \textit{structured reference string (SRS)}) between them in order to execute the proof and verification. 
    Because a leak of this secret value allows fake provers to generate a valid proof, it must be destroyed at the end of this setup phase to keep the protocol secure. Because of this potential danger, it is called \textit{toxic waste}. Some (not all) non-interactive proof algorithms require this step. If a protocol does not require a \textit{trusted setup}, it is called a \textit{transparent} protocol.
	
	\item \textit{Single} vs. \textit{universal} setup: if a \textit{trusted setup} is required per circuit (i.e., a graph that represent mathematical operations), it is called \textit{single setup}. If a \textit{trusted setup} can be used for all circuits, it is called \textit{universal setup}. 
\end{itemize} 

\begin{figure*}[!htbp]
	\centering
    \centerline{\includegraphics[width=18cm]{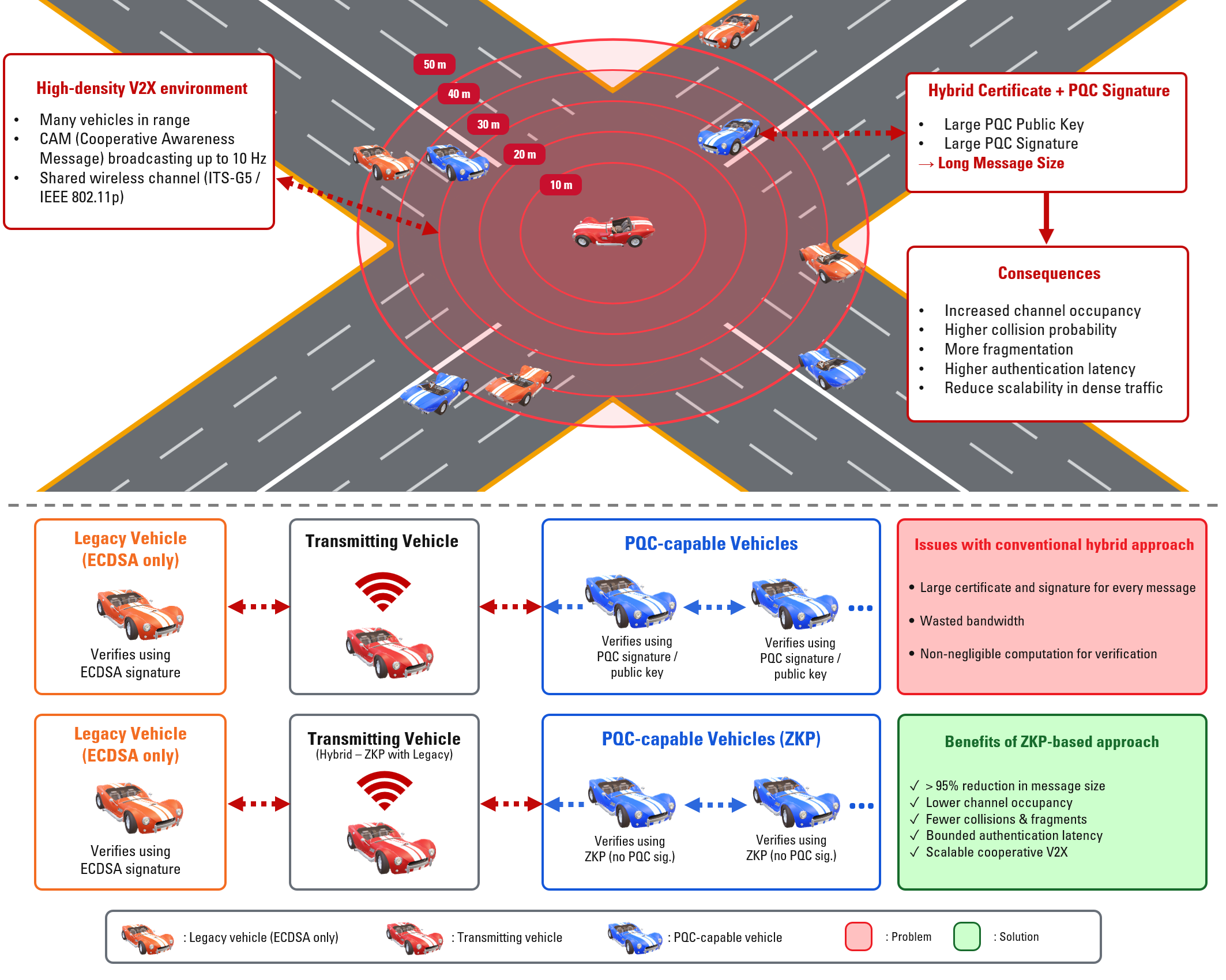}}
	\caption{ZKS-PQC Concept}
	\label{fig:zkspqc_concept}
\end{figure*}

Protocols in zk-SNARKs, as the name implies, are \textit{succinct} and \textit{non-interactive}. 
They have the characteristics of a fixed proof length and non-interactive, implying that a \textit{trusted setup} is required. However, not all require it.

Earlier ZKP protocols are interactive. However, more recent protocols are non-interactive. 
This characteristic is especially suitable for distributed systems and asynchronous verification, such as blockchains. 
Table~\ref{tab:zkp_overview} shows the high-level overview of ZKP protocols.

Bulletproofs is transparent (i.e., it does not require a trusted setup), have a logarithmic-size proof ($O(log(n))$), and is non-interactive zero-knowledge proof based on inner-product arguments.
It is especially efficient for range proofs $0\leq x<2^n$ without revealing $x$.
It also performs short zero-knowledge proofs for
general arithmetic circuits~\cite{bunz2018bulletproofs}.

\section{Proposed Scheme} \label{sec:proposed_scheme}
\subsection{ZKS-PQC Concept}
To minimize the increase in message size caused by the introduction of PQC DSA, we propose a new scheme called ZKS-PQC.
Applying shorter fixed-size ZKP proof and commitment as the signature verification replaces a longer signature or public key of the PQC DSA.
This way, the increase in message size is kept significantly lower compared to PQC signature algorithms that have long signatures or public keys.
This overall concept is presented in Fig.~\ref{fig:zkspqc_concept}.

\begin{figure}[!htbp]
	\centering
	\centerline{\includegraphics[width=7.5cm]{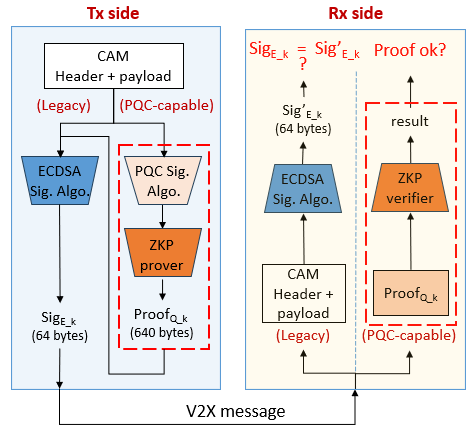}}
	\caption{ZKS-PQC Processing}
	\label{fig:zkspqc_processing}
\end{figure}

Details processing of the ZKP-PQC scheme is illustrated in Fig.~\ref{fig:zkspqc_processing}.
Instead of the conventional signature verification, in which receivers calculate an expected signature and compare it against the received signature, receivers verify the ZKP proof of the signature the transmitter generates.
Therefore, messages no longer need to include both the public key and the signature of the PQC DSA.
As a result, the overall messag size becomes constant regardless of the PQC DSA used.

The introduction of ZKP is done in a backward-compatible manner so that it is transparent to vehicles that support the legacy ECDSA only.
In other words, vehicles that support PQC DSA generate both the PQC DSA and ECDSA signatures to ensure a smooth migration.

\begin{figure*}[!htbp]
	\centering
	\centerline{\includegraphics[width=14cm]{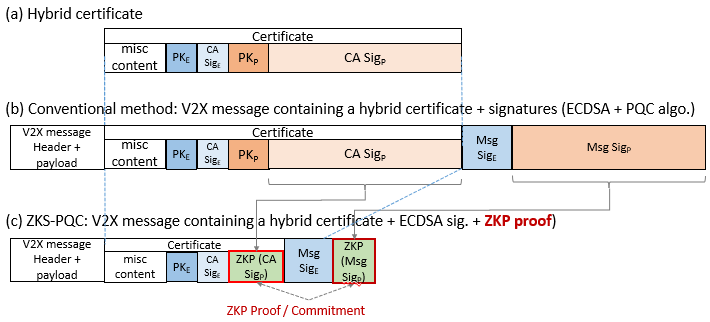}}
	\caption{Message Structure}
	\label{fig:message_structure}
\end{figure*}

The resulting message format is shown in Fig.~\ref{fig:message_structure}. 
In this figure, (a) represents the hybrid certificate in which both ECDSA and PQC public keys and CA signatures are present.\footnote{The CA signatures are issued by the issuing CA and verify the authenticity of the certificate.} 
Also, (b) shows the conventional message format where message signatures generated by both ECDSA and PQC algorithms and PQC-DSA public key are present.
Contrary to (b), (c) shows our scheme, where the ZKP proof/commitment replaces the PQC-based (significantly longer) signature and public key.\footnote{The suffix ``E'' and ``P'' in the information elements represent the ECDSA-based and the PQC-based elements, respectively.} 

\subsection{Definition}
In this section, we present a formal definition of our scheme and pseudocode.

\textbf{Definition 1.} Let $$\mathcal{S}_i=(\mathsf{Sign_P}_i(), \mathsf{Prf}_i(), \mathsf{Extr}_i(), \mathsf{Vrfy_P}_i(), \mathsf{Sign_E}_i(), \mathsf{Vrfy_E}_i())$$ be the ZKS-PQC scheme for all $i\in\mathbb{N}$, $m$ be the upper-layer message ($m\in\mathcal{M}$), where $\mathcal{M}$ is the message space of all V2X messages.
We define the V2X protocol $\mathcal{P}=(\mathsf{MsgGen(), MsgVrfy()})$ based on the following algorithms:\footnote{For brevity, we omit the formal definition of the function to extract the ZKP proof portion from the received message $msg$, as it is a rather peripheral part of this definition.} 
\begin{itemize}
	\item $\mathsf{MsgGen(}m\mathsf{)}$ for a given upper-layer message $m$ consists of four steps: (1) calculate a signature for $m$ based on the PQC DSA of choice, (2) generate a ZKP proof for the PQC signature, (3) calculate a ECDSA signature, and (4) concatenate the message, the certificate, the ZKP proof, and the ECDSA signature, i.e., $\mathsf{MsgGen(}m\mathsf{)}=m\mathbin\Vert cert \mathbin\Vert\mathsf{Prf(Sign_P(}m\mathsf{)})\mathbin\Vert\mathsf{Sign_E(}m\mathsf{)}$.
	Thus, $msg \leftarrow \mathsf{MsgGen(}m\mathsf{)}$
    
	\item $\mathsf{MsgVrfy(}msg\mathsf{)}$ consist of two steps: (1) if the receiver is PQC-capable, extract the ZKP proof from the received V2X message $msg$, and verify the proof, i.e., $\mathsf{MsgVrfy(}msg\mathsf{)}=\mathsf{Vrfy_P(Extr(}msg\mathsf{)})$, (2) if the receiver supports legacy ECDSA only, then verify the ECDSA signature, i.e., $\mathsf{MsgVrfy(}msg\mathsf{)}=\mathsf{Vrfy_E(}msg\mathsf{)})$.
    The result of $\mathsf{MsgVrfy()}$ is either \textit{true} (valid) or \textit{false} (invalid). Thus, $1|0 \leftarrow \mathsf{MsgVrfy(}msg\mathsf{)}$. 
\end{itemize}

Algorithms for $\mathsf{MsgGen()}$ and $\mathsf{MsgVrfy()}$ are respectively illustrated in Algorithm~\ref{fig:alg_tx} and~\ref{fig:alg_rx}.

\begin{algorithm}[]
    \footnotesize
	\SetAlgoLined
	\KwResult{generate PQC/ECDSA hybrid mode message}
    $m \leftarrow$ upper layer message\;
    $Prf_P \leftarrow \mathsf{Prf(Sign_P(}m\mathsf{))}$\;
    $Sig_E \leftarrow \mathsf{Sign_E(}m\mathsf{))}$\;
    $msg \leftarrow m\mathbin\Vert cert\mathbin\Vert Prf_P \mathbin\Vert Sig_E$\; 
    Send($msg$)\;
	\caption{$\mathsf{MsgGen()}$ message transmission}
	\label{fig:alg_tx}
\end{algorithm}

\begin{algorithm}[]
    \footnotesize
	\SetAlgoLined
	\KwResult{verify PQC/ECDSA hybrid mode message}
    $msg \leftarrow Rcv()$\;
    \If{(PQC-capable)}{
        \tcp{$\mathsf{Vrfy_P()}$ part}
        $prf \leftarrow \mathsf{Extr(}msg\mathsf{)}$\;
    	$res \leftarrow \mathsf{Vrfy_P(}prf\mathsf{))}$\;
        \If{($res=True$)}{
            accept message\;
        }
        \Else{
            reject message\;
        }
    }
    \Else{
        (conventional ECDSA signature verification)\; 
        }
	\caption{$\mathsf{MsgVrfy()}$ message verification}
	\label{fig:alg_rx}
\end{algorithm}

\section{Experiment} \label{sec:experiment}
We conducted an experiment to verify our ZKS-PQC scheme and measured its performance. 
For our purpose, we selected Bulletproofs~\cite{bunz2018bulletproofs} as the ZKP algorithm.
As discussed in Sect.~\ref{sec:zkpprimer}, it is a \textit{non-interactive} protocol and no \textit{trusted setup} is required.
These characteristics are requirements in V2X communication, given that its basic services are one-way broadcast messages among moving vehicles where they encounter a constantly changing random number of unknown vehicles on the road.
Therefore, V2X messages are transmitted without prior interactions with other vehicles, and the proof needs to be self-sufficient for verification.

\subsection{Methodology}
\begin{figure}[!htbp]
	\centering
	\centerline{\includegraphics[width=7cm]{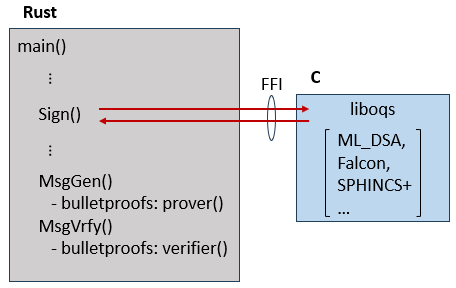}}
	\caption{ZKS-PQC Implementation}
	\label{fig:implementation}
\end{figure}

We used two open source implementations: open-quantum safe (OQS) written in C~\cite{githubliboqs} and Bulletproofs written in Rust~\cite{githubbulletproofs}. 
Due to this circumstance, it required a binding function to build an executable object. 
The main routine, including Bulletproofs, is written in Rust, and it calls the C OQS library for signature algorithms, as shown in Fig.~\ref{fig:implementation}. 
The \textit{release mode} compile option in Rust was used to build the executable object for optimal performance.\footnote{The default mode of Rust compilation includes debug-related information and makes the resulting object significantly slower than the optimized \textit{release} mode (\textit{'RUSTFLAGS="-C target-cpu=native" cargo run --release'}).}

We implemented all variants of all three standardized PQC algorithms (ML-DSA (i.e., Dilithium), Falcon, and SPHINCS+) and some of the algorithms from the ongoing evaluation in NIST (CROSS, FAEST, UOV, MAYO).\footnote{CROSS is now eliminated from further consideration according to NIST's round 2 report~\cite{nistIr8610}.}
For each variant of each algorithm, we ran the sender and receiver operations 1000 times and took an average value.
Performance results include the time to perform the operations on both the sender and receiver sides: \textit{prover time}, the time to generate a signature and its proof, and \textit{verifier time}, the time for the verifier to verify the proof.

We ran these algorithms on two different hardware. 
The first one is a generic laptop equipped with an Intel Core i7-6600 CPU@2.60GHz processor with 16GB of DDR4 memory, running Ubuntu Linux version 22.04.
The second is a development platform for C-V2X, called MK6C, by Cohda Wireless.

The MK6C module is equipped with Qualcomm's MDM 9150 chipset, and its OS is based on Ubuntu Linux. 
It can be used as either OBU or RSU, depending on the configuration to run the code.
It supports multiple connectivity, such as CAN, Ethernet, WiFi, and Bluetooth.
Because the binary object built in the Intel-based laptop is not compatible with the Qualcomm processor's architecture, we rebuilt the object on MK6C by repeating the same build process to compile the same Rust-based code to execute the object on it.\footnote{We were unable to successfully compile code for FAEST on MK6C even after multiple efforts. Therefore, we do not have the performance numbers for the FAEST-Bulletproofs combination, as shown by the blank space for FAEST in the right half of Table~\ref{tab:performance}.}

\subsection{Simulation Results}
Table~\ref{tab:msg_size_summary} and~\ref{tab:performance} capture our simulation results - message size reduction ratio and the performance summary of various PQC signature algorithms combined with ZKP Bulletproofs to realize our ZKS-PQC concept as shown in Fig.~\ref{fig:zkspqc_processing}.

\begin{table*}[!htbp]
	\centering
	\caption{Message Size Comparison of Conventional vs. ZKS-PQC}
	\label{tab:msg_size_summary}
	\begin{center}
		\begin{tabular}{
				>{\raggedright\arraybackslash}p{0.2\textwidth}
				>{\centering\arraybackslash}p{0.04\textwidth}|
				>{\raggedleft\arraybackslash}p{0.06\textwidth}
				>{\raggedleft\arraybackslash}p{0.05\textwidth}
				>{\raggedleft\arraybackslash}p{0.06\textwidth}|
				>{\raggedleft\arraybackslash}p{0.06\textwidth}
				>{\raggedleft\arraybackslash}p{0.06\textwidth}|
				>{\raggedleft\arraybackslash}p{0.06\textwidth}
				>{\raggedleft\arraybackslash}p{0.06\textwidth}|
				>{\raggedleft\arraybackslash}p{0.07\textwidth}
			}
			\noalign{\hrule height 1pt}
			~ & ~ & (a) & (b) & (c) & (d) & (e) & (f) & (g) & (h) \\ \hline
			\multicolumn{1}{c}{\textbf{Algorithm}} 
			& \textbf{Security Level} 
			& \textbf{$|$PK$|$ (byte)} 
			& \textbf{$|$Sig$|$ (byte)} 
			& \textbf{$|$Cert$|$ (byte) (note 1)} 
			& \textbf{$|$Hybrid Cert\_conv$|$  (byte) (note 2) }
			& \textbf{{$|$CAM\_ conv$|$ (byte) (note 3)}} 
			& \textbf{$|$Hybrid Cert\_ zks-sig$|$  (byte) (note 4)} 
			& \textbf{$|$CAM\_ zks-sig$|$ (byte) (note 5)} 
			& \textbf{Msg size reduction (1-(g)/(e))} \\ 
			\noalign{\hrule height 1pt}
			ECDSA & -- & 32 & 64 & 150 & -- & -- & -- & -- & -- \\ \hline
			ML\_DSA\_44 & 2 & 1312 & 2420 & 3786 & 3882 & 6516 & 790 & 1644 & 74.770\% \\ 
			ML\_DSA\_65 & 3 & 1952 & 3309 & 5315 & 5411 & 8934 & 790 & 1644 & 81.598\% \\ 
			ML\_DSA\_87 & 5 & 2592 & 4627 & 7273 & 7369 & 12210 & 790 & 1644 & 86.536\% \\ \hline
			Falcon-512 & 1 & 897 & 752 & 1703 & 1799 & 2765 & 790 & 1644 & 40.542\% \\ 
			Falcon-1024 & 5 & 1793 & 1462 & 3309 & 3405 & 5081 & 790 & 1644 & 67.644\% \\ \hline
			DILITHIUM2 & 2 & 1312 & 2420 & 3786 & 3882 & 6516 & 790 & 1644 & 74.770\% \\ 
			DILITHIUM3 & 3 & 1952 & 3293 & 5299 & 5395 & 8902 & 790 & 1644 & 81.532\% \\ 
			DILITHIUM5 & 5 & 2592 & 4595 & 7241 & 7337 & 12146 & 790 & 1644 & 86.465\% \\ \hline
			SPHINCS\_SHA2\_128f\_simple & 1 & 32 & 17088 & 17174 & 17270 & 34572 & 790 & 1644 & 95.245\% \\ 
			SPHINCS\_SHA2\_128s\_simple & 1 & 32 & 7856 & 7942 & 8038 & 16108 & 790 & 1644 & 89.794\% \\ 
			SPHINCS\_SHA2\_192f\_simple & 3 & 48 & 35664 & 35766 & 35862 & 71740 & 790 & 1644 & 97.708\% \\ 
			SPHINCS\_SHA2\_192s\_simple & 3 & 48 & 16224 & 16326 & 16422 & 32860 & 790 & 1644 & 94.997\% \\ 
			SPHINCS\_SHA2\_256f\_simple & 5 & 64 & 49856 & 49974 & 50070 & 100140 & 790 & 1644 & 98.358\% \\ 
			SPHINCS\_SHA2\_256s\_simple & 5 & 64 & 29792 & 29910 & 30006 & 60012 & 790 & 1644 & 97.261\% \\ 
			SPHINCS\_SHAKE\_128f\_simple & 1 & 32 & 17088 & 17174 & 17270 & 34572 & 790 & 1644 & 95.245\% \\ 
			SPHINCS\_SHAKE\_128s\_simple & 1 & 32 & 7856 & 7942 & 8038 & 16108 & 790 & 1644 & 89.794\% \\ 
			SPHINCS\_SHAKE\_192f\_simple & 3 & 48 & 35664 & 35766 & 35862 & 71740 & 790 & 1644 & 97.708\% \\ 
			SPHINCS\_SHAKE\_192s\_simple & 3 & 48 & 16224 & 16326 & 16422 & 32860 & 790 & 1644 & 94.997\% \\ 
			SPHINCS\_SHAKE\_256f\_simple & 5 & 64 & 49856 & 49974 & 50070 & 100140 & 790 & 1644 & 98.358\% \\ 
			SPHINCS\_SHAKE\_256s\_simple & 5 & 64 & 29792 & 29910 & 30006 & 60012 & 790 & 1644 & 97.261\% \\ \hline
			CROSS\_RSDP\_128\_fast & 1 & 77 & 18432 & 18563 & 18659 & 37305 & 790 & 1644 & 95.593\% \\ 
			CROSS\_RSDP\_128\_small & 1 & 77 & 12432 & 12563 & 12659 & 25305 & 790 & 1644 & 93.503\% \\ 
			CROSS\_RSDP\_128\_balanced & 1 & 77 & 13152 & 13283 & 13379 & 26745 & 790 & 1644 & 93.853\% \\ 
			CROSS\_RSDP\_192\_fast & 3 & 115 & 41406 & 41575 & 41671 & 83291 & 790 & 1644 & 98.026\% \\ 
			CROSS\_RSDP\_192\_small & 3 & 115 & 28391 & 28560 & 28656 & 57261 & 790 & 1644 & 97.129\% \\ 
			CROSS\_RSDP\_192\_balanced & 3 & 115 & 29853 & 30022 & 30118 & 60185 & 790 & 1644 & 97.268\% \\ 
			CROSS\_RSDP\_256\_fast & 5 & 153 & 74590 & 74797 & 74893 & 149697 & 790 & 1644 & 98.902\% \\ 
			CROSS\_RSDP\_256\_small & 5 & 153 & 50818 & 51025 & 51121 & 102153 & 790 & 1644 & 98.391\% \\ 
			CROSS\_RSDP\_256\_balanced & 5 & 153 & 53527 & 53734 & 53830 & 107571 & 790 & 1644 & 98.472\% \\ 
			CROSS\_RSDPG\_128\_fast & 1 & 54 & 11980 & 12088 & 12184 & 24378 & 790 & 1644 & 93.256\% \\ 
			CROSS\_RSDPG\_128\_small & 1 & 54 & 8960 & 9068 & 9164 & 18338 & 790 & 1644 & 91.035\% \\ 
			CROSS\_RSDPG\_128\_balanced & 1 & 54 & 9120 & 9228 & 9324 & 18658 & 790 & 1644 & 91.189\% \\ 
			CROSS\_RSDPG\_192\_fast & 3 & 83 & 26772 & 26909 & 27005 & 53991 & 790 & 1644 & 96.955\% \\ 
			CROSS\_RSDPG\_192\_small & 3 & 83 & 20452 & 20589 & 20685 & 41351 & 790 & 1644 & 96.024\% \\ 
			CROSS\_RSDPG\_192\_balanced & 3 & 83 & 22464 & 22601 & 22697 & 45375 & 790 & 1644 & 96.377\% \\ 
			CROSS\_RSDPG\_256\_fast & 5 & 106 & 48102 & 48262 & 48358 & 96674 & 790 & 1644 & 98.299\% \\ 
			CROSS\_RSDPG\_256\_small & 5 & 106 & 36454 & 36614 & 36710 & 73378 & 790 & 1644 & 97.760\% \\ 
			CROSS\_RSDPG\_256\_balanced & 5 & 106 & 40100 & 40260 & 40356 & 80670 & 790 & 1644 & 97.962\% \\ \hline
			FAEST\_128f & 1 & 32 & 5924 & 6010 & 6106 & 12244 & 790 & 1644 & 86.573\% \\ 
			FAEST\_128s & 1 & 32 & 4506 & 4592 & 4688 & 9408 & 790 & 1644 & 82.526\% \\ 
			FAEST\_192f & 3 & 48 & 14948 & 15050 & 15146 & 30308 & 790 & 1644 & 94.576\% \\ 
			FAEST\_192s & 3 & 48 & 11260 & 11362 & 11458 & 22932 & 790 & 1644 & 92.831\% \\ 
			FAEST\_256f & 5 & 48 & 26548 & 26650 & 26746 & 53508 & 790 & 1644 & 96.928\% \\ 
			FAEST\_256s & 5 & 48 & 20696 & 20798 & 20894 & 41804 & 790 & 1644 & 96.067\% \\ \hline
			MAYO\_1 & 1 & 1420 & 454 & 1928 & 2024 & 2692 & 790 & 1644 & 38.930\% \\ 
			MAYO\_2 & 2 & 4912 & 186 & 5152 & 5248 & 5648 & 790 & 1644 & 70.892\% \\ 
			MAYO\_3 & 3 & 2986 & 681 & 3721 & 3817 & 4712 & 790 & 1644 & 65.110\% \\ 
			MAYO\_5 & 5 & 5554 & 964 & 6572 & 6668 & 7846 & 790 & 1644 & 79.047\% \\ \hline
			OV\_Is & 1 & 412160 & 96 & 412310 & 412406 & 412716 & 790 & 1644 & 99.602\% \\ 
			OV\_Ip & 1 & 278432 & 128 & 278614 & 278710 & 279052 & 790 & 1644 & 99.411\% \\ 
			OV\_III & 3 & 1225440 & 200 & 1225694 & 1225790 & 1226204 & 790 & 1644 & 99.866\% \\ 
			OV\_V & 5 & 2869440 & 260 & 2869754 & 2869850 & 2870324 & 790 & 1644 & 99.943\% \\ 
			OV\_Is\_pkc & 1 & 66576 & 96 & 66726 & 66822 & 67132 & 790 & 1644 & 97.551\% \\ 
			OV\_Ip\_pkc & 1 & 43576 & 128 & 43758 & 43854 & 44196 & 790 & 1644 & 96.280\% \\ 
			OV\_III\_pkc & 3 & 189232 & 200 & 189486 & 189582 & 189996 & 790 & 1644 & 99.135\% \\ 
			OV\_V\_pkc & 5 & 446992 & 260 & 447306 & 447402 & 447876 & 790 & 1644 & 99.633\% \\ 
			OV\_Is\_pkc\_skc & 1 & 66576 & 96 & 66726 & 66822 & 67132 & 790 & 1644 & 97.551\% \\ 
			OV\_Ip\_pkc\_skc & 1 & 43576 & 128 & 43758 & 43854 & 44196 & 790 & 1644 & 96.280\% \\ 
			OV\_III\_pkc\_skc & 3 & 189232 & 200 & 189486 & 189582 & 189996 & 790 & 1644 & 99.135\% \\ 
			OV\_V\_pkc\_skc & 5 & 446992 & 260 & 447306 & 447402 & 447876 & 790 & 1644 & 99.633\% \\ 
			\noalign{\hrule height 1pt}
			\multicolumn{10}{l}{\begin{tabular}[c]{@{}l@{}}
					Note 1: (PQC pub. key + sig.) + other cert. elements.\\ 
					Note 2: (PQC pub. key + sig.) + (ECDSA pub. key + sig.) + other cert. elements.\\
					Note 3: hybrid cert.(d) + (PQC sig. + ECDSA sig.) + CAM msg. header/payload.\\
					Note 4: (ECDSA pub. key + sig.) + (ZKP proof + commt.) + other cert. elements.\\
					Note 5: hybrid cert.(f) + ECDSA sig. + (ZKP proof + commt.) + CAM msg. header/payload.
			\end{tabular}}
		\end{tabular}
	\end{center}
\end{table*}

\begin{table*}[!htbp]
	\centering
	\caption{Performance Summary}
	\label{tab:performance}
	\begin{center}
		\begin{tabular}{l|c|rrrr|rrrr}
			\noalign{\hrule height 1pt}
			\multirow{2}{*}{Algorithm} & \multirow{2}{*}{\begin{tabular}[c]{@{}l@{}}Sec.\\ Lvl\end{tabular}} & \multicolumn{4}{c|}{Linux Laptop}    & \multicolumn{4}{c}{CW MK6C}      \\ \cline{3-10} 
			&      & \multicolumn{1}{l}{\begin{tabular}[c]{@{}l@{}}PQC \\ sign\\ (msec)\end{tabular}} & \multicolumn{1}{l|}{\begin{tabular}[c]{@{}l@{}}PQC \\ sig. \\ verify\\ (msec)\end{tabular}} & \multicolumn{1}{l}{\begin{tabular}[c]{@{}l@{}}BP \\ prover\\ (msec)\end{tabular}} & \begin{tabular}[c]{@{}l@{}}BP\\ verifier\\ (msec)\end{tabular} & \multicolumn{1}{l}{\begin{tabular}[c]{@{}l@{}}PQC\\ sign\\ (msec)\end{tabular}} & \multicolumn{1}{l|}{\begin{tabular}[c]{@{}l@{}}PQC\\ sig.\\ Verify\\ (msec)\end{tabular}} & \multicolumn{1}{l}{\begin{tabular}[c]{@{}l@{}}BP\\ prover\\ (msec)\end{tabular}} & \begin{tabular}[c]{@{}l@{}}BP\\ verifier\\ (msec)\end{tabular} \\ 
			\noalign{\hrule height 1pt}
			
			Falcon-512  & 1 & \multicolumn{1}{r}{0.003}  & \multicolumn{1}{r|}{$<$0.001} & \multicolumn{1}{r}{19.259} & 2.010 & \multicolumn{1}{r}{2.000} & \multicolumn{1}{r|}{$<$0.001} & \multicolumn{1}{r}{81.099} & 11.000 \\ 
			
			Falcon-1024 & 5  & \multicolumn{1}{r}{1.004} & \multicolumn{1}{r|}{$<$0.001} & \multicolumn{1}{r}{19.526} & 2.009 & \multicolumn{1}{r}{4.000} & \multicolumn{1}{l|}{$<$0.001} & \multicolumn{1}{r}{83.091} & 11.000 \\ \hline
			
			ML\_DSA\_44 & 2  & \multicolumn{1}{r}{0.001} & \multicolumn{1}{r|}{$<$0.001} & \multicolumn{1}{r}{18.349} & 2.009 & \multicolumn{1}{r}{3.641} & \multicolumn{1}{r|}{1.000} & \multicolumn{1}{r}{83.458} & 11.002 \\ 
			
			ML\_DSA\_65 & 3  & \multicolumn{1}{r}{0.010} & \multicolumn{1}{r|}{$<$0.001} & \multicolumn{1}{r}{18.579} & 2.017 & \multicolumn{1}{r}{6.527} & \multicolumn{1}{r|}{1.000} & \multicolumn{1}{r}{86.155} & 11.000 \\ 
			
			ML\_DSA\_87 & 5  & \multicolumn{1}{r}{0.031} & \multicolumn{1}{r|}{$<$0.001} & \multicolumn{1}{r}{19.045} & 2.022 & \multicolumn{1}{r}{7.698} & \multicolumn{1}{r|}{2.000} & \multicolumn{1}{r}{87.327} & 11.014 \\ \hline
			
			SPHINCS\_SHA2\_128f\_simple & 1 & \multicolumn{1}{r}{22.927} & \multicolumn{1}{r|}{1.507} & \multicolumn{1}{r}{41.653} & 2.050 & \multicolumn{1}{r}{228.389} & \multicolumn{1}{r|}{13.148} & \multicolumn{1}{r}{307.956} & 11.000 \\ 
			
			SPHINCS\_SHA2\_128s\_simple & 1 & \multicolumn{1}{r}{480.027} & \multicolumn{1}{r|}{0.015} & \multicolumn{1}{r}{498.533} & 2.027 & \multicolumn{1}{r}{4880.338} & \multicolumn{1}{r|}{4.211} & \multicolumn{1}{r}{4959.746} & 11.000 \\ 
			
			SPHINCS\_SHA2\_192f\_simple & 3 & \multicolumn{1}{r}{39.861} & \multicolumn{1}{r|}{2.116} & \multicolumn{1}{r}{58.304}  & 2.035 & \multicolumn{1}{r}{399.473} & \multicolumn{1}{r|}{20.903} & \multicolumn{1}{r}{479.073} & 11.001 \\ 
			
			SPHINCS\_SHA2\_192s\_simple & 3 & \multicolumn{1}{r}{903.407} & \multicolumn{1}{r|}{1.005} & \multicolumn{1}{r}{921.735} & 2.028 & \multicolumn{1}{r}{9243.848} & \multicolumn{1}{r|}{7.089} & \multicolumn{1}{r}{9323.383} & 11.003 \\ 
			
			SPHINCS\_SHA2\_256f\_simple & 5 & \multicolumn{1}{r}{84.298} & \multicolumn{1}{r|}{2.402} & \multicolumn{1}{r}{102.840} & 2.038 & \multicolumn{1}{r}{820.892} & \multicolumn{1}{r|}{21.263} & \multicolumn{1}{r}{900.315} & 11.000 \\ 
			
			SPHINCS\_SHA2\_256s\_simple & 5 & \multicolumn{1}{r}{840.575} & \multicolumn{1}{r|}{1.056} & \multicolumn{1}{r}{859.257} & 2.083 & \multicolumn{1}{r}{8485.135} & \multicolumn{1}{r|}{10.902} & \multicolumn{1}{r}{8564.591} & 11.001 \\ 
			
			SPHINCS\_SHAKE\_128f\_simple & 1 & \multicolumn{1}{r}{45.265} & \multicolumn{1}{r|}{3.003} & \multicolumn{1}{r}{63.425} & 2.020 & \multicolumn{1}{r}{447.022} & \multicolumn{1}{r|}{26.122} & \multicolumn{1}{r}{526.626} & 11.000 \\ 
			
			SPHINCS\_SHAKE\_128s\_simple & 1 & \multicolumn{1}{r}{944.075} & \multicolumn{1}{r|}{1.001} & \multicolumn{1}{r}{962.436} & 2.031 & \multicolumn{1}{r}{9270.618} & \multicolumn{1}{r|}{8.680} & \multicolumn{1}{r}{9350.079} & 11.001 \\ 
			
			SPHINCS\_SHAKE\_192f\_simple & 3 & \multicolumn{1}{r}{73.483} & \multicolumn{1}{r|}{4.031} & \multicolumn{1}{r}{91.892}  & 2.030 & \multicolumn{1}{r}{733.907} & \multicolumn{1}{r|}{38.925} & \multicolumn{1}{r}{813.035} & 11.000 \\ 
			
			SPHINCS\_SHAKE\_192s\_simple & 3 & \multicolumn{1}{r}{1630.537} & \multicolumn{1}{r|}{1.021} & \multicolumn{1}{r}{1649.014} & 2.042 & \multicolumn{1}{r}{16091.952} & \multicolumn{1}{r|}{12.887} & \multicolumn{1}{r}{16171.388} & 11.001 \\ 
			
			SPHINCS\_SHAKE\_256f\_simple & 5 & \multicolumn{1}{r}{151.421} & \multicolumn{1}{r|}{4.118} & \multicolumn{1}{r}{169.882}  & 2.043 & \multicolumn{1}{r}{1484.882} & \multicolumn{1}{r|}{39.147} & \multicolumn{1}{r}{1564.401} & 11.001 \\ 
			
			SPHINCS\_SHAKE\_256s\_simple & 5 & \multicolumn{1}{r}{1429.957} & \multicolumn{1}{r|}{2.015} & \multicolumn{1}{r}{1448.392} & 2.022 & \multicolumn{1}{r}{14309.391} & \multicolumn{1}{r|}{19.134} & \multicolumn{1}{r}{14388.781} & 11.000 \\ \hline
			
			CROSS\_RSDP\_128\_fast & 1 & \multicolumn{1}{r}{1.002} & \multicolumn{1}{r|}{0.002}  & \multicolumn{1}{r}{19.275} & 2.020 & \multicolumn{1}{r}{6.001} & \multicolumn{1}{r|}{3.007} & \multicolumn{1}{r}{85.986} & 11.000 \\ 
			
			CROSS\_RSDP\_128\_small & 1 & \multicolumn{1}{r}{4.006} & \multicolumn{1}{r|}{2.315} & \multicolumn{1}{r}{22.522} & 2.017 & \multicolumn{1}{r}{25.000} & \multicolumn{1}{r|}{14.000} & \multicolumn{1}{r}{104.776} & 11.000 \\ 
			
			CROSS\_RSDP\_128\_balanced & 1 & \multicolumn{1}{r}{2.003} & \multicolumn{1}{r|}{1.002}  & \multicolumn{1}{r}{21.048} & 2.031 & \multicolumn{1}{r}{12.000} & \multicolumn{1}{r|}{7.000} & \multicolumn{1}{r}{91.922} & 11.000 \\ 
			
			CROSS\_RSDP\_192\_fast & 3 & \multicolumn{1}{r}{2.079} & \multicolumn{1}{r|}{1.014} & \multicolumn{1}{r}{21.353} & 2.029 & \multicolumn{1}{r}{18.000} & \multicolumn{1}{r|}{10.000} & \multicolumn{1}{r}{97.875} & 11.000 \\ 
			
			CROSS\_RSDP\_192\_small & 3 & \multicolumn{1}{r}{6.632} & \multicolumn{1}{r|}{4.192} & \multicolumn{1}{r}{25.812} & 2.028 & \multicolumn{1}{r}{47.611} & \multicolumn{1}{r|}{22.000} & \multicolumn{1}{r}{127.285} & 11.022 \\ 
			
			CROSS\_RSDP\_192\_balanced & 3 & \multicolumn{1}{r}{4.051} & \multicolumn{1}{r|}{3.000}  & \multicolumn{1}{r}{23.264} & 2.025 & \multicolumn{1}{r}{32.000} & \multicolumn{1}{r|}{15.000} & \multicolumn{1}{r}{111.183} & 11.003 \\ 
			
			CROSS\_RSDP\_256\_fast & 5 & \multicolumn{1}{r}{4.570} & \multicolumn{1}{r|}{3.004} & \multicolumn{1}{r}{23.479} & 2.018 & \multicolumn{1}{r}{31.000} & \multicolumn{1}{r|}{18.000} & \multicolumn{1}{r}{110.998} & 11.000 \\ 
			
			CROSS\_RSDP\_256\_small & 5 & \multicolumn{1}{r}{12.881} & \multicolumn{1}{r|}{8.065}  & \multicolumn{1}{r}{31.805} & 2.016 & \multicolumn{1}{r}{86.002} & \multicolumn{1}{r|}{39.489} & \multicolumn{1}{r}{165.913} & 11.000 \\ 
			
			CROSS\_RSDP\_256\_balanced & 5 & \multicolumn{1}{r}{7.760} & \multicolumn{1}{r|}{5.008}  & \multicolumn{1}{r}{26.967} & 2.023 & \multicolumn{1}{r}{53.000} & \multicolumn{1}{r|}{25.469} & \multicolumn{1}{r}{132.143} & 11.000 \\ 
			
			CROSS\_RSDPG\_128\_fast & 1 & \multicolumn{1}{r}{0.627} & \multicolumn{1}{r|}{0.011}  & \multicolumn{1}{r}{21.188} & 2.400 & \multicolumn{1}{r}{5.000} & \multicolumn{1}{r|}{3.000} & \multicolumn{1}{r}{84.601} & 11.000 \\ 
			
			CROSS\_RSDPG\_128\_small & 1 & \multicolumn{1}{r}{3.004} & \multicolumn{1}{r|}{2.002} & \multicolumn{1}{r}{21.180} & 2.009 & \multicolumn{1}{r}{23.000} & \multicolumn{1}{r|}{14.000} & \multicolumn{1}{r}{102.034} & 11.000 \\ 
			
			CROSS\_RSDPG\_128\_balanced & 1 & \multicolumn{1}{r}{1.011} & \multicolumn{1}{r|}{1.000}  & \multicolumn{1}{r}{19.418} & 2.012 & \multicolumn{1}{r}{11.000} & \multicolumn{1}{r|}{7.000} & \multicolumn{1}{r}{90.771} & 11.000 \\ 
			
			CROSS\_RSDPG\_192\_fast & 3 & \multicolumn{1}{r}{1.751} & \multicolumn{1}{r|}{1.018} & \multicolumn{1}{r}{20.901} & 2.096 & \multicolumn{1}{r}{12.000} & \multicolumn{1}{r|}{7.000} & \multicolumn{1}{r}{91.004} & 11.000 \\ 
			
			CROSS\_RSDPG\_192\_small & 3 & \multicolumn{1}{r}{4.526} & \multicolumn{1}{r|}{3.006}  & \multicolumn{1}{r}{23.426} & 2.012 & \multicolumn{1}{r}{30.645} & \multicolumn{1}{r|}{18.997} & \multicolumn{1}{r}{110.166} & 11.006 \\ 
			
			CROSS\_RSDPG\_192\_balanced & 3 & \multicolumn{1}{r}{2.014} & \multicolumn{1}{r|}{1.009} & \multicolumn{1}{r}{21.374} & 2.023 & \multicolumn{1}{r}{16.000} & \multicolumn{1}{r|}{10.000} & \multicolumn{1}{r}{95.318} & 11.006 \\ 
			
			CROSS\_RSDPG\_256\_fast & 5 & \multicolumn{1}{r}{3.005} & \multicolumn{1}{r|}{2.001}  & \multicolumn{1}{r}{22.148} & 2.016 & \multicolumn{1}{r}{21.798} & \multicolumn{1}{r|}{14.000} & \multicolumn{1}{r}{101.060} & 11.001 \\ 
			
			CROSS\_RSDPG\_256\_small & 5 & \multicolumn{1}{r}{7.023} & \multicolumn{1}{r|}{4.422} & \multicolumn{1}{r}{25.889} & 2.024 & \multicolumn{1}{r}{51.000} & \multicolumn{1}{r|}{30.000} & \multicolumn{1}{r}{130.013} & 11.000 \\ 
			
			CROSS\_RSDPG\_256\_balanced & 5 & \multicolumn{1}{r}{4.000} & \multicolumn{1}{r|}{2.011} & \multicolumn{1}{r}{22.837} & 2.010 & \multicolumn{1}{r}{28.000} & \multicolumn{1}{r|}{17.000} & \multicolumn{1}{r}{107.045} & 11.000 \\ \hline
			
			FAEST\_128f & 1 & \multicolumn{1}{r}{9.172} & \multicolumn{1}{r|}{7.376} & \multicolumn{1}{r}{32.668} & 3.181  & \multicolumn{1}{l}{}    & \multicolumn{1}{l|}{}   & \multicolumn{1}{l}{}  &    \\ 
			
			FAEST\_128s & 1 & \multicolumn{1}{r}{63.021} & \multicolumn{1}{r|}{45.376} & \multicolumn{1}{r}{85.530} & 2.942 & \multicolumn{1}{l}{}    & \multicolumn{1}{l|}{}   & \multicolumn{1}{l}{}  &    \\ 
			
			FAEST\_192f & 3 & \multicolumn{1}{r}{23.457} & \multicolumn{1}{r|}{18.946} & \multicolumn{1}{r}{45.813} & 2.992 & \multicolumn{1}{c}{See}    & \multicolumn{1}{c|}{note}   & \multicolumn{1}{c}{See}  & note   \\ 
			
			FAEST\_192s & 3 & \multicolumn{1}{r}{211.209} & \multicolumn{1}{r|}{122.597} & \multicolumn{1}{r}{234.547} & 3.120 & \multicolumn{1}{l}{}    & \multicolumn{1}{l|}{}   & \multicolumn{1}{l}{}  &    \\ 
			
			FAEST\_256f & 5 & \multicolumn{1}{r}{36.651} & \multicolumn{1}{r|}{33.134} & \multicolumn{1}{r}{58.752}  & 2.963 & \multicolumn{1}{l}{}    & \multicolumn{1}{l|}{}   & \multicolumn{1}{l}{}  &    \\ 
			
			FAEST\_256f & 5 & \multicolumn{1}{r}{267.579} & \multicolumn{1}{r|}{257.934} & \multicolumn{1}{r}{292.185} & 3.330 & \multicolumn{1}{l}{}    & \multicolumn{1}{l|}{}   & \multicolumn{1}{l}{}  &    \\ \hline
			
			MAYO\_1 & 1 & \multicolumn{1}{r}{$<$0.001} & \multicolumn{1}{r|}{$<$0.001} & \multicolumn{1}{r}{18.232} & 2.009 & \multicolumn{1}{r}{3.000} & \multicolumn{1}{r|}{1.000} & \multicolumn{1}{r}{82.008} & 11.000 \\ 
			
			MAYO\_2 & 2 & \multicolumn{1}{r}{$<$0.001} & \multicolumn{1}{r|}{$<$0.001} & \multicolumn{1}{r}{18.343} & 2.018 & \multicolumn{1}{r}{2.000} & \multicolumn{1}{l|}{$<$0.001} & \multicolumn{1}{r}{81.003} & 11.000 \\ 
			
			MAYO\_3 & 3 & \multicolumn{1}{r}{1.001} & \multicolumn{1}{r|}{$<$0.001} & \multicolumn{1}{r}{19.697} & 2.024 & \multicolumn{1}{r}{8.674} & \multicolumn{1}{r|}{4.000} & \multicolumn{1}{r}{87.999} & 11.000 \\ 
			
			MAYO\_5 & 5 & \multicolumn{1}{r}{2.712} & \multicolumn{1}{r|}{1.051} & \multicolumn{1}{r}{23.535} & 2.337 & \multicolumn{1}{r}{21.992} & \multicolumn{1}{r|}{9.000} & \multicolumn{1}{r}{101.032} & 11.000 \\ \hline
			
			OV\_Is & 1 & \multicolumn{1}{r}{$<$0.001}  & \multicolumn{1}{r|}{$<$0.001} & \multicolumn{1}{r}{18.981} & 2.157 & \multicolumn{1}{r}{1.069} & \multicolumn{1}{r|}{$<$0.001} & \multicolumn{1}{r}{80.547} & 11.033 \\ 
			
			OV\_Ip & 1 & \multicolumn{1}{r}{$<$0.001}  & \multicolumn{1}{r|}{$<$0.001} & \multicolumn{1}{r}{19.259} & 2.159 & \multicolumn{1}{r}{0.004} & \multicolumn{1}{r|}{$<$0.001} & \multicolumn{1}{r}{79.999} & 11.005 \\ 
			
			OV\_III & 3 & \multicolumn{1}{r}{$<$0.001} & \multicolumn{1}{r|}{$<$0.001} & \multicolumn{1}{r}{20.250} & 2.284 & \multicolumn{1}{r}{3.008} & \multicolumn{1}{r|}{1.000} & \multicolumn{1}{r}{82.050} & 11.001 \\ 
			
			OV\_V & 5 & \multicolumn{1}{r}{0.006} & \multicolumn{1}{r|}{$<$0.001} & \multicolumn{1}{r}{18.946} & 2.013 & \multicolumn{1}{r}{7.047} & \multicolumn{1}{r|}{2.000} & \multicolumn{1}{r}{86.989} & 11.000 \\ 
			
			OV\_Is\_pkc & 1 & \multicolumn{1}{r}{$<$0.001}  & \multicolumn{1}{r|}{$<$0.001} & \multicolumn{1}{r}{18.141} & 2.010 & \multicolumn{1}{r}{1.055} & \multicolumn{1}{r|}{1.876} & \multicolumn{1}{r}{80.103} & 11.001 \\ 
			
			OV\_Ip\_pkc & 1 & \multicolumn{1}{r}{$<$0.001}  & \multicolumn{1}{r|}{$<$0.001} & \multicolumn{1}{r}{18.068} & 2.015 & \multicolumn{1}{r}{0.010}  & \multicolumn{1}{r|}{1.000} & \multicolumn{1}{r}{80.018} & 11.012 \\ 
			
			OV\_III\_pkc & 3 & \multicolumn{1}{r}{$<$0.001} & \multicolumn{1}{r|}{0.114} & \multicolumn{1}{r}{18.092} & 2.007 & \multicolumn{1}{r}{3.024}  & \multicolumn{1}{r|}{5.001} & \multicolumn{1}{r}{82.159} & 11.054 \\ 
			
			OV\_V\_pkc & 5 & \multicolumn{1}{r}{0.004} & \multicolumn{1}{r|}{2.004} & \multicolumn{1}{r}{18.345} & 2.011 & \multicolumn{1}{r}{7.024}  & \multicolumn{1}{r|}{11.003} & \multicolumn{1}{r}{86.991} & 11.000   \\ 
			
			OV\_Is\_pkc\_skc & 1 & \multicolumn{1}{r}{2.001} & \multicolumn{1}{r|}{$<$0.001} & \multicolumn{1}{r}{20.181} & 2.007 & \multicolumn{1}{r}{22.599} & \multicolumn{1}{r|}{1.975} & \multicolumn{1}{r}{102.046} & 11.000 \\ 
			
			OV\_Ip\_pkc\_skc & 1 & \multicolumn{1}{r}{1.002} & \multicolumn{1}{r|}{$<$0.001} & \multicolumn{1}{r}{19.258} & 2.009 & \multicolumn{1}{r}{9.000} & \multicolumn{1}{r|}{1.000}  & \multicolumn{1}{r}{88.023} & 11.000 \\ 
			
			OV\_III\_pkc\_skc & 3 & \multicolumn{1}{r}{8.009} & \multicolumn{1}{r|}{0.089} & \multicolumn{1}{r}{26.144} & 2.008 & \multicolumn{1}{r}{60.012} & \multicolumn{1}{r|}{5.000} & \multicolumn{1}{r}{139.629} & 11.000 \\ 
			
			OV\_V\_pkc\_skc & 5 & \multicolumn{1}{r}{18.229} & \multicolumn{1}{r|}{2.000} & \multicolumn{1}{r}{36.510} & 2.011 & \multicolumn{1}{r}{163.379} & \multicolumn{1}{r|}{11.043} & \multicolumn{1}{r}{242.895} & 11.036 \\ 
			
			\noalign{\hrule height 1pt}
            \multicolumn{10}{l}{Note: no data due to unsuccessful compilation attempts.}
		\end{tabular}
	\end{center}
\end{table*}

\subsubsection{Message Size Reduction} \label{sec:msg_size_red}
\begin{figure}[!htbp]%
\centering {{\includegraphics[width=8.5cm]{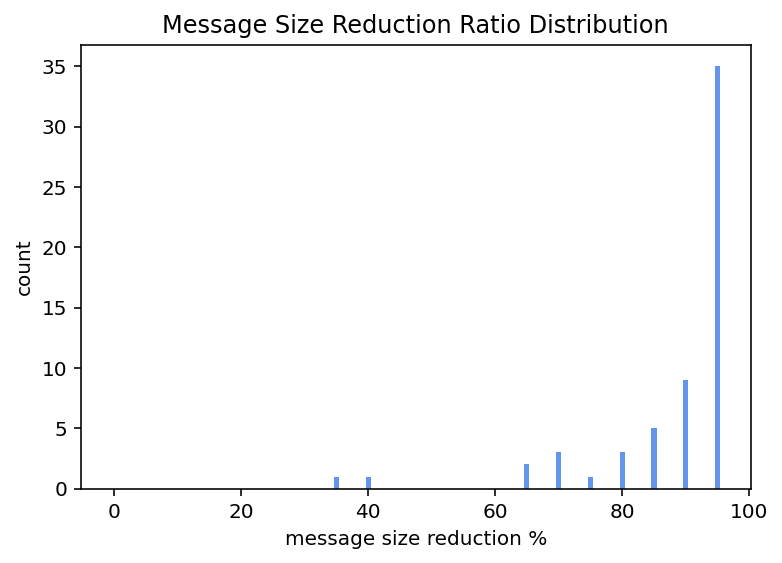} }}%
	\caption{Message Size Reduction}%
	\label{fig:msg_size_reduction}%
\end{figure}

Table~\ref{tab:msg_size_summary} shows the message size reduction ratio as a result of ZKS-PQC scheme (Fig.~\ref{fig:message_structure}(c)) compared against the conventional message format, where it contains the PQC signature and PQC public key in the certificate, (Fig.~\ref{fig:message_structure}(b)). 
The \% number for each algorithm experimented is shown in the right-most column in Table~\ref{tab:msg_size_summary}, and 
Figure~\ref{fig:msg_size_reduction} captures its high-level summary by counting the number of algorithms based on the range of message size reduction ratio in increments of 5\%.

As shown in Fig.~\ref{fig:msg_size_reduction}, the level of improvement is significant, with over 95\% message size reduction in the large proportion of the algorithms. 
This includes both: (1) algorithms that have a large signature size and a small public key size, such as SPHINCS+, CROSS, FAEST, and (2) algorithms that have a small signature size and a large public key size, such as MAYO and UOV (multivariate-based DSAs).
Both ML-DSA (Dilithium) and Falcon have less reduction ratio at between 40 to 86\%.
This reflects their relatively smaller signature/public key size compared to other DSAs.

\subsubsection{Sign/Prover and Verifier Performance}
\begin{figure*}[!htbp]%
	\centering
	\subfloat[\centering PQC Sign Gen. Time] {{\includegraphics[width=8.5cm]{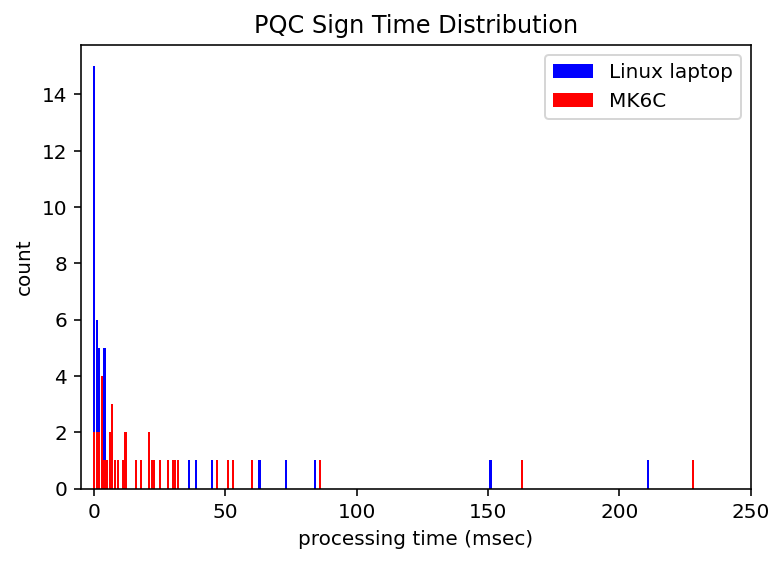} }}%
	\qquad
	\subfloat[\centering PQC Sig. Verification Time] {{\includegraphics[width=8.5cm]{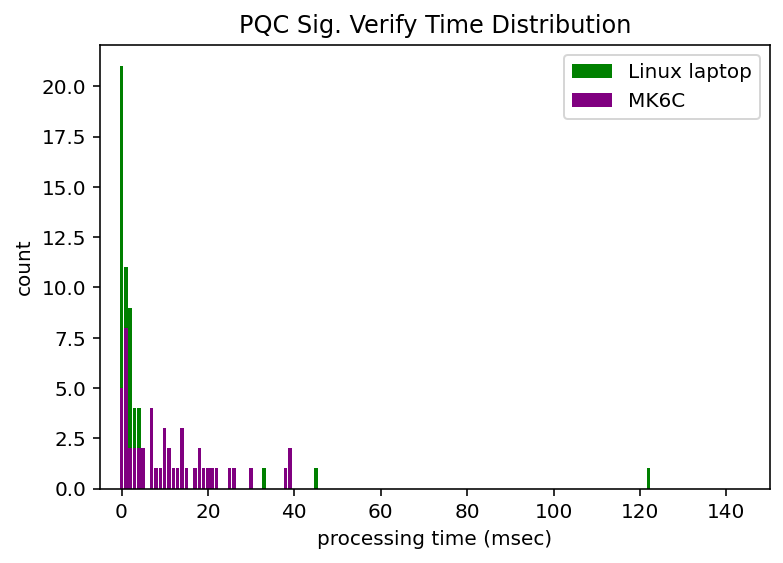} }}%
	\caption{PQC Sig. Gen/Verify Performance}%
	\label{fig:PQC_sig_verify_performance}%
\end{figure*}

\begin{figure*}[!htbp]%
	\centering
	\subfloat[\centering BP Prover Time] {{\includegraphics[width=8cm]{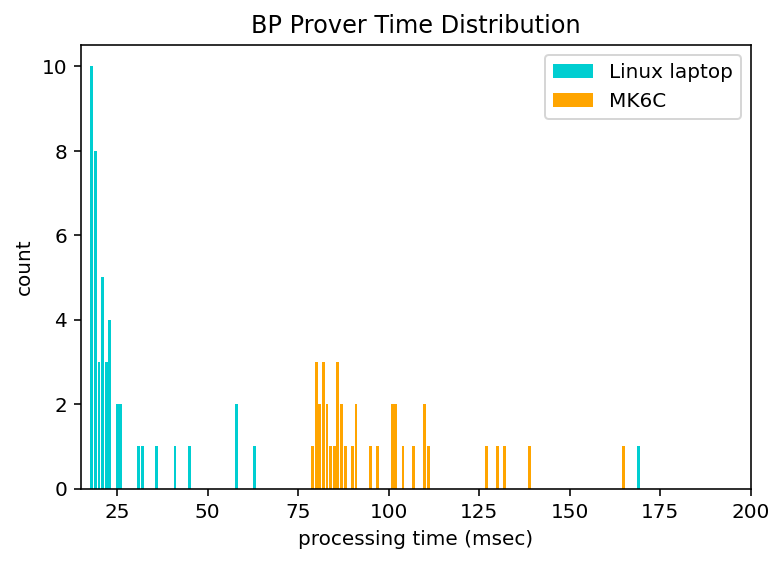} }}%
	\qquad
	\subfloat[\centering BP Verifier Time] {{\includegraphics[width=8cm]{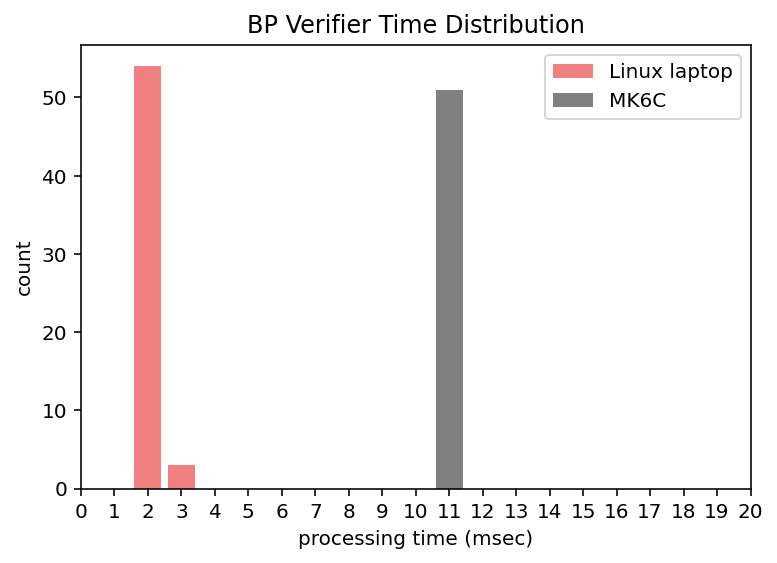} }}%
	\caption{BP ZKP Prover/Verifier Performance}%
	\label{fig:BP_proof_gen_verify_performance}%
\end{figure*}

Table~\ref{tab:performance} shows the prover and verifier time in milliseconds.
The third column to the sixth column show the result for the Linux laptop, and the seventh to the tenth column show the corresponding numbers for MK6C.
Another view of this table in visual representation for the PQC signature and verification for both Linux laptop and MK6C are captured in  
Fig.~\ref{fig:PQC_sig_verify_performance}.(a) and (b).
Similarly, Bulletproofs prover and verifier for both Linux laptop and MK6C are shown in  
Fig.~\ref{fig:BP_proof_gen_verify_performance}.(a) and (b).

From the PQC signature generation and verification perspectives, there are wide variations between these two hardware platforms.
Specifically, slower algorithms, such as SPHINCS, exhibit more pronounced difference between them as MK6C is almost 10 times slower than Linux laptop in both PQC signature generation/verification and Bulletproofs proof generation/verification processing.
For relatively faster algorithms, such as ML-DSA and CROSS, MK6C is approximately 3 to 6 times slower.

From the Bulletproofs prover and verifier perspectives, these two platforms show more pronounced performance differences between them.
Specifically, the processing delay in MK6C compared to the Linux laptop is clearly visible in Figure~\ref{fig:BP_proof_gen_verify_performance}.(a) and (b).

The key observation is that: (1) the PQC signature generation and verification in Linux laptop are consistently fast around several msec and 4 to 8 times slower in MK6C, (2) the bulletproofs prover time for the majority of algorithms concentrates between 18 and 23 msec for Linux laptop and 4 to 8 times slower in MK6C, and (3) the verifier time across all algorithms concentrates at around 2 msec for Linux laptop, and 11 msec for MK6C.\footnote{The difference in the prover time directly reflects the signature generation of the algorithms and their variants.}
The exceptions are slower variants of SPHINCS+ (designated with ``s'' in their names) where the prover time is between sub-1 sec. to 1.6 sec for the Linux laptop and approximately 10 times slower in MK6C, proportional to the security level. 


\subsection{Evaluation and Analysis}
Our scheme makes the message size constant irrespective of the algorithm chosen, as a result of removing both public key and signature and replacing them with a fixed-size proof and commitment.
In fact, the effectiveness of the message size reduction becomes progressively pronounced as the combined size of signature and public key increases. 
The processing overhead of our scheme stays at a reasonable level (18$\sim$23 msec) for the prover and negligible time ($\sim$2 msec) for the verifier across all algorithms in Linux laptop. 
In other words, our scheme neutralizes the impact of message size increase and levels out the performance overhead across the range of PQC signature algorithms.
As a result, our scheme \textit{effectively brings all PQC DSAs to the same playing field} and allows algorithm selection without being concerned with the impacts of additional message size increase with reasonable performance level.
This means that a higher security level of any algorithm can be selected in RTS and communication systems without significant performance degradation and channel usage increase. 
It is beneficial for real-time communication systems, such as V2X, where the minimum message interval of CAM messages is 100 msec. 
We consider this a significant point, since the conventional message structure (Fig.~\ref{fig:message_structure}.(b)) will make the message length prohibitively long.

The relative lower performance of MK6C raises a concern as it is slower than a generic Linux laptop across all algorithms, specifically in SPHINCS, which is approximately 10 times slower.
We cannot draw a conclusion on this point just by experimenting with one development platform from a single vendor.
However, it possibly hints at practical migration constraints toward PQC algorithms in the future if similar limitations apply to other commercial OBU hardware platforms in vehicles.

\subsection{Impacts on Automated Driving Systems Communications} \label{sec:ads}
The measured message-size reductions decrease the authentication material carried across the communication interface connecting CAV connectivity services with the vehicle-control and backend assets identified in the EU coordinated risk assessment on connected and automated vehicles~\cite
{NISEUcoordinated2026}. 

The ZKS-PQC scheme retains ECDSA verification for legacy receivers while allowing compatible receivers to verify the zero-knowledge proof. 
It therefore supports mixed-fleet migration and complements the certificate architecture of ETSI TS 103 097~\cite{etsiTs1030972022}.

This authentication layer precedes the use of V2X data by the CAV sensor stack. Although sensor fusion improves robustness against spoofed CAM, delayed CAM verification may be combined with attacks against LiDAR or camera inputs~\cite{farrell2023security}. 
In this regard, ZKS-PQC provides a foundation for quantum-resilient CAV authentication by addressing authentication-material size and processing cost, but it does not establish the trustworthiness of downstream sensor fusion or vehicle behavior.

The framework may also apply to other authenticated services, including event-driven Decentralized Environmental Notification Messages (DENM) and quasi-periodic Collective Perception Messages (CPM). CPM communicate sensor capabilities, perceived objects, and perception regions from vehicles or roadside RSUs~\cite{ts103324v2112024}. 
These services have different payloads and generation patterns and therefore require separate evaluation. 
The present Bulletproofs implementation provides a proof-of-concept baseline; future iterations require integration of a post-quantum-secure proof system and evaluation under dense-network and operational ADS conditions.

\subsection{Additional Considerations on ZKP Protocol} \label{sec:zkp}
Our selection of Bulletproofs as the ZKP algorithm in our experiment is to demonstrate the effectiveness of our scheme.
We note that Bulletproofs is based on discrete logarithm problem~\cite{bunz2018bulletproofs} and it is one of the mathematical problems that are expected to be broken by Shor's algorithm using sufficiently powerful quantum computers.
Therefore, it is not post-quantum ready protocol by definition.
Ideally, it should be replaced with another post-quantum resistant ZKP protocol.
There are multiple such ZKP protocols that meets the requirements (i.e., non-interactive, transparent, and universal) to be used in the V2X communication systems. 
They include Virgo~\cite{zhang2020transparent}, Ligero~\cite{ames2022ligero}, Aurora~\cite{ben2019aurora}, zk-STARK~\cite{ben2019scalable}, and Zilch~\cite{mouris2021zilch}. 
However, none of these protocols' performance appears to meet the requirements of real-time systems (i.e., fast proof and verification processing). 
This is a gap that needs to be filled. 
The discussion in~\cite{bunz2018bulletproofsiacr} hints required modifications to make Bulletproofs post-quantum resistant. 
We consider addressing this point to be our next step.

\section{Conclusion} \label{sec:conclusion}
In this paper, we introduced ZKS-PQC - a novel approach to replace large signatures and public keys, which is a characteristic of many PQC DSAs, with compact ZKP proofs.
Through our experiment, we demonstrated that our scheme neutralizes the impact of message size increase incurred by many PQC DSAs, and the additional processing overhead for Bulletproofs is within an acceptable level in real-time communication systems, such as V2X, and likely in other RTS systems also.
As a result, our scheme broadens the portfolio of usable PQC DSAs for RTS and communication systems.
Due to its backward-compatible design, our scheme fully supports a seamless and incremental transition in the migration period from ECDSA to PQC DSAs.
The present Bulletproofs implementation establishes an architectural and experimental baseline of our scheme.
Our next step is to experiment integrating PQC DSAs with a post-quantum-secure ZKP protocol, and evaluate additional V2X services and network-level operating conditions for further improvements, and validate the scheme on multiple hardware platforms.

\section*{Acknowledgment}
This work was supported by the Flemish Government through the Cybersecurity Research Program with grant number: VOEWICS02.

For the preparation of this paper, we used Grammarly for spell checking, grammatical error correction, and text improvement.

\bibliographystyle{IEEEtran}
\bibliography{bibliography}

\end{document}